\documentclass[aip, amsmath,amssymb, reprint]{revtex4-1}

\usepackage{graphicx}
\usepackage{dcolumn}
\usepackage{bm}

\usepackage{amssymb}
\usepackage{amsmath}
\usepackage{times}
\usepackage{xcolor} 
\usepackage{babel}[English]

\usepackage[normalem]{ulem}

\makeatletter
\def\@email#1#2{%
 \endgroup
 \patchcmd{\titleblock@produce}
  {\frontmatter@RRAPformat}
  {\frontmatter@RRAPformat{\produce@RRAP{*#1\href{mailto:#2}{#2}}}\frontmatter@RRAPformat}
  {}{}
}%
\makeatother
\begin{document}

\title[Dynamics and chaotic properties of the fully disordered Kuramoto model]
 {Dynamics and chaotic properties of the fully disordered Kuramoto model}


\author{Iv\'an Le\'on}
\affiliation{Departamento de Matemática Aplicada y Ciencias de la Computación, Universidad de Cantabria, Santander, Spain,}
\author{Diego Paz\'o}
\affiliation{Instituto de Física de Cantabria (IFCA), Universidad de Cantabria–CSIC, 39005 Santander, Spain}

\date{\today}

\begin{abstract}

Frustrated random interactions are a key ingredient of spin glasses.
From this perspective,
we study the dynamics of
the Kuramoto model with quenched random couplings: the simplest oscillator ensemble
with fully disordered interactions.
We answer some open questions by means of extensive numerical simulations and a
perturbative calculation (the cavity method).
We show frequency entrainment is not realized in the thermodynamic limit and
that chaotic dynamics are pervasive in parameter space.
In the weak coupling regime, we find closed formulas for the frequency shift and
the dissipativeness of the model.
Interestingly, the largest Lyapunov exponent is found to
exhibit the same asymptotic dependence on the coupling constant
irrespective of the coupling asymmetry,  within the numerical accuracy.

\end{abstract}

  \maketitle

\begin{quotation}

A wide variety of natural and artificial systems are modeled as populations of
interacting limit-cycle oscillators\cite{PRK01}.
Studying the emergent phenomena in these systems
is a well-established research field.
For simplicity, it is usual to consider one-dimensional (phase) oscillators since they capture the dynamics of limit-cycle attractors, provided the interactions are sufficiently weak\cite{Kur84}.
In this context, a prominent system is the
Kuramoto model\cite{Kur75}, describing the transition to collective synchronization of  heterogeneous phase oscillators
under all-to-all attractive 
interactions.
However, this is only the tip of the iceberg,
as many other phenomena arise for other interaction schemes.


A particularly interesting situation arises when 
 attractive and repulsive interactions 
are randomly spread in the population. 
In such a situation a complex behavior can be foreseen
due to the so-called frustration,
as occurs in
spin-glass models \cite{SK75}.
Indeed, with the purpose of observing an `oscillator glass'
in 1992 Daido \cite{Dai92} modified
the
Kuramoto model introducing random frustrated interactions.
Arguably, his numerical simulations
displayed a quasi-glassy phase.
A few years later, Stiller and Radons \cite{StiRad98}
generalized the model allowing the interactions to be nonsymmetric.
Chaotic dynamics was shown to be preponderant in that case.
Analytical descriptions, borrowing techniques from spin glass theory \cite{HU08,Kimo19,Pruser24a,Pruser24b},
and numerical studies \cite{Pazo23,Nic25} have appeared in recent years.
In spite of these advances, our understanding of the model remains quite limited.

In this paper we revisit the problem answering open questions and improving previous results. 
We resort to extensive numerical simulations, and the perturbative cavity method
recently developed for this system \cite{Pruser24a,Pruser24b}. 
Our numerical simulations indicate that in the thermodynamic limit
($N\to\infty$),
oscillators do not get entrained for any finite value of the parameters. Furthermore, we show the pervasiveness of chaotic dynamics in the thermodynamic limit.
 Specifically, for small coupling, 
the value of the largest Lyapunov exponent 
 appears to be insensitive to the
asymmetry of the coupling.

\end{quotation}
  
\section{Introduction} \label{sec.intro}

Synchronization and other emergent phenomena in populations of oscillators
have been profusely studied due to their relevance in many fields,
from biology to engineering \cite{PRK01}.
The Kuramoto model\cite{Kur75,Kur84} stood out as a simple system capable of describing
the transition to collective synchrony with minimal ingredients.
In particular, the interactions between the (phase) oscillators were chosen
global, uniform, and pairwise,
and the only source of disorder is provided by the heterogeneous natural frequencies.
Along the years, these simplifying assumptions have been relaxed in different ways:
including more sophisticated forms of disorder
\cite{MP11,PM11,MP11p,IMS14,PikBagn24,SmirnovPik24} or setting complex coupling topologies,  like random networks \cite{rodriguez16,chiba19}
or the actual human-brain connectome network \cite{villegas14}.

Decades ago, motivated by the complex behavior of spin glasses,
a modified Kuramoto model
with fully disordered interactions
(of Sherrington-Kirkpatrick type \cite{SK75})
was put forward by Daido \cite{Dai92}.
He reported the existence of an `oscillator glass', a state
displaying  ``quasientrainment'' \cite{Dai92}, as well as
slow (algebraic) relaxation to the asymptotic state \cite{Dai92,Dai00,SR00}.
In Ref.~\onlinecite{Dai92}, it was argued that such quasi-glassy state
was concomitant with a concave up density of local fields at zero;
a ``volcano state'' in the recent jargon \cite{Ottino18}.
In recent works the volcano transition was studied
adopting a low-rank coupling matrix \cite{Ottino18,Pazo23}. Borrowing techniques from spin-glass theory, the genuine full-rank coupling matrix was studied\cite{HU08,Kimo19} and led
to a rigorous analysis of the volcano transition by means of a perturbative approach\cite{Pruser24a,Pruser24b}.

The model proposed by Daido assumed symmetric (i.e.~reciprocal) interactions. This
restriction, grounded in physical laws for magnetic interactions,
is unrealistic in many other domains.
Indeed, the effect of nonreciprocity has
been investigated in several disordered systems.
While some systems \cite{Spitzner89,EissOpper94}
show robust spin-glass features,
in other systems the spin-glass
phase is fragile \cite{CrisSom87}.
Stiller and Radons
analyzed the Daido model with asymmetric interactions \cite{StiRad98},
concluding that chaos was generic in the model.
(Incidentally, chaos is also observed in similar models
of coupled oscillators with phase-lag disorder \cite{PikBagn24}.)
However, Stiller and Radons also reported that a
region with non-chaotic behavior, denoted as ``freezing'',
survived for weak nonreciprocity and large enough coupling.

In this paper, we revisit the Daido-Stiller-Radons
(DSR) model, in an attempt to clarify
certain questions: Is quasi-entrainment (frequency entrainment without phase locking) real?
Which is the exact domain of chaos? Is non-chaotic behavior robust against weak non-reciprocalities and heterogeneity? Our work provides answers to these questions, and at the same time, it points to an
apparent universal scaling of chaos in the weak coupling regime.
Moreover, we obtain an asymptotic law for the dissipativeness of the model
in the weak coupling regime.

This paper is organized as follows. In section \ref{sec.model} we introduce the studied model. We provide a brief survey of the observed phenomenology
and customary methodologies in Sec.~\ref{sec.background}.
Sections \ref{sec.weakcoup} and \ref{sec.infcoup} are devoted to the study of the dynamics and Lyapunov exponents on two special limits of the model, weak and infinitely strong coupling. Then we complete the picture in Sec.~\ref{sec.modcoup} by analyzing the moderate and strong coupling regimes. Finally, we present the conclusions in Sec.~\ref{sec.conc}.

\section{Model} \label{sec.model}

In this paper we consider the Kuramoto model with quenched random
interactions or Daido-Stiller-Radons (DSR) model, for short.
It consists of $N$ oscillators,
with their phases $\{\theta_j\}_{j=1,\ldots,N}$ governed by the differential equations:
\begin{equation}\label{eq.phasmod}
	\frac{d \theta_j}{dt}\equiv\dot{\theta}_j=\omega_j + \frac{J}{\sqrt{N}}\sum_{k=1}^{N}K_{jk}\sin(\theta_k-\theta_j) .
\end{equation}
The intrinsic frequencies $\{\omega_j\}$ are independently drawn from
a normal distribution $g(\omega)$ with
variance $\sigma^2$  and zero mean (by going to a rotating frame, if necessary).
The heterogeneous couplings are encoded in the
matrix elements $K_{jk}$, drawn from a zero-mean unit-variance normal distribution
(diagonal elements vanish: $K_{jj}=0$).
The correlation between diagonally opposed matrix elements is:
\begin{equation}
	\langle K_{jk}K_{kj}\rangle =\eta \qquad \mbox{($j\ne k$)} .
	\label{K}
\end{equation}
The parameter $\eta\in[-1,1]$
controls the reciprocity of the coupling. The specific values
$\eta=1$, 0 and $-1$ correspond to
reciprocal, fully asymmetric, and anti-reciprocal coupling, respectively.
The dynamics of the model \eqref{eq.phasmod}
also depends on the coupling constant $J$.
We have, in sum, a set of ODEs
with $N(N-1)$ quenched random coefficients $K_{jk}$ and $N$
random natural frequencies $\omega_j$. For large enough $N$
we can focus on ``typical'' behaviors controlled by three continuous parameters:
$\sigma$, $J$ and $\eta$.

For future discussion, we cast system \eqref{eq.phasmod}
with each oscillator driven by its own local field:
\begin{subequations}
 \begin{eqnarray}
	\dot{\theta}_j&=&\omega_j +  J \, r_j \sin(\psi_j-\theta_j)  , \label{eq.lf} \\
    r_j e^{i\psi_j}&=&\frac1{\sqrt N} \sum_{k=1}^N K_{jk} e^{i\theta_k} \label{eq.LF}
 \end{eqnarray}
\end{subequations}

It is important to notice that
only the ratio between $\sigma$ and $J$ matters.
Introducing a rescaled time $\tau=J t$,
and expressing the natural frequencies as
$\omega=\sigma\tilde{\omega}$, with $\tilde\omega_j\sim{\cal N}(0,1)$,
the model becomes:
\begin{equation}\label{eq.phasmodrescaled}
	\frac{d{\theta}_j}{d\tau}=\frac{\sigma}{J}\tilde{\omega}_j + \frac{1}{\sqrt{N}}\sum_{k=1}^{N}K_{jk}\sin(\theta_k-\theta_j) .
\end{equation}
Written in this form it is clear that there are only
two relevant continuous parameters in the model:
the rescaled frequency dispersion $\sigma/J$, and the asymmetry of the coupling $\eta$.
Nevertheless, it is important to keep in mind
the other ingredients:
the population size $N$, the particular realization
of the disorder, and the initial condition for the phases.

It will be also convenient to refer to an alternative representation of the model:
\begin{equation}\label{eq.phasmodrescaled2}
	\frac{d{\theta}_j}{dT}=\tilde{\omega}_j
	+ \frac{J/\sigma}{\sqrt{N}}\sum_{k=1}^{N}K_{jk}\sin(\theta_k-\theta_j) .
\end{equation}
Now the rescaled time is $T=\sigma t$.


\section{Background} \label{sec.background}


Once the DSR model has been properly introduced, it is
convenient to review its main properties.
We will first focus on its phenomenology.
The specific methodologies to tackle the model are afterwards introduced.

\subsection{Phenomenology}
In the original works by Daido\cite{Dai92}
and Stiller and Radons \cite{StiRad98} certain novel phenomena arising
in Eq.~\eqref{eq.phasmod} were described.
In what follows we briefly comment on these phenomena and the current understanding
of them.

\subsubsection{Glassy phase at zero temperature}

To contextualize the phenomenology of the DSR model, we first revisit
the limiting case of identical oscillators
($\sigma=0$) and reciprocal interactions ($\eta=1$, $J\ne0$).
In this particular case
the phases obey gradient dynamics:
\begin{equation}
	\dot\theta_j=-\frac{\partial V}{\partial \theta_j} ,
\end{equation}
with potential

\begin{equation}
	V({\bm \theta})
	=-\frac{J}{\sqrt{N}} \sum_{j=1}^{N-1}\sum_{k=j+1}^N K_{jk} \cos(\theta_k-\theta_j) .
\end{equation}

In turn, the system eventually settles into a resting state,
i.e.~a stable fixed point,
among an exponential number (with $N$) of them due to
the frustrated interactions
\cite{MPV87}.
The relaxation process of the global order parameter
\begin{equation}
|Z|=\frac1N\left|\sum_{k}^Ne^{i\theta_k}\right|
\label{Z}
\end{equation}
turns out to be ``superslow'' ($\sim1/\ln t$), according to the
numerical results in Ref.~\onlinecite{Dai18}.

\subsubsection{Volcano transition}
The question whether the just
described glassy phase is robust for nonzero $\sigma$
(in analogy to the persistence of magnetic spin-glasses
at finite temperature) motivated Daido\cite{Dai92} to introduce and
study the DSR model (with $\eta=1$).
It is important to notice that for nonzero $\sigma$ the
DSR model loses its gradient character.
Daido found that, above a critical coupling $J_v$,
the density $\rho(r)$ of the local-field amplitudes,
see Eq.~\eqref{eq.LF}, becomes concave up at zero.
(Recent numerical estimations of the critical coupling
are $J_v\approx1.3\sigma$ \cite{Pazo23,Pruser24a}.)
Hence the local fields organize like a volcano in the complex plane.
Daido argued the volcano phase was concomitant with
the quasientrainment between the oscillators,
and the algebraic relaxation of $|Z|$
(see below).

\subsubsection{Slow relaxation}
The algebraic decay to $0$ of the global order parameter \eqref{Z}
was presented in Ref.~\onlinecite{Dai92} as a signature of a ``quasi-glassy'' state
in  the volcano phase.
This fact drew some  controversy \cite{StiRad98,Dai00,SR00}, in part because
the average over coupling matrices can be inside
or outside the absolute value.
Pr\"user et al.\cite{Pruser24a} recently argued
that the volcano transition
cannot be linked to the glassy transition. However,
recent numerical simulations with $N=10^4$ oscillators\cite{Nic25}
support the algebraic decay of $|Z|$ in the volcano phase.
If the onset of this algebraic decay coincides with $J=J_v$
deserves further scrutiny since,
according to Ref.~\onlinecite{Nic25},
it shows up only if the coupling
is progressively increased with $N$ as $J\sim\sqrt{N}$.

\subsubsection{Quasientrainment}
Another phenomenon observed in the volcano phase was the quasientrainment of oscillators \cite{Dai92}. It consisted
in an apparent frequency locking of
some oscillators, but with their phases diffusing (i.e., no phase-locking).
Despite the puzzling character of this behavior
in a deterministic system, it remains insufficiently
explored in our opinion.

\subsubsection{Freezing}
Stiller and Radons observed a quasi-glassy state in the form of a frozen state
for strong couplings, $J>J_f\approx24$ (far beyond the volcano transition).
Actually, the value of $\sigma$ was not specified in Ref.~\onlinecite{StiRad98},
but we can presume $ \sigma=1$.
Freezing was described as a state in which all oscillators became entrained to the central frequency.
However, only small population sizes ($N=100-400$) were studied,
so this result might not hold in the thermodynamic limit.

\subsubsection{The effect of nonreciprocity}

The robustness of the previous phenomena against coupling asymmetry ($\eta<1$)
is an interesting question.
Stiller and Radons \cite{StiRad98}
concluded that a quasi-glassy
(freezing) phase
indeed persisted for $\eta_c(J)<\eta<1$, with $\lim_{J\to\infty}\eta_c(J)\approx0.8$. Again, this conclusion is supported by simulations with
a small number of oscillators ($N=100$), see Fig.~8 in Ref.~\onlinecite{StiRad98}.

Concerning the volcano phase, it was  recently shown
that it persists for non-reciprocal coupling\cite{Pazo23}.
Specifically, the volcano exists in the range
$\eta>\eta_v\approx0.2$. This result
is consistent with a recent perturbative calculation,
valid for small $J$\cite{Pruser24b}.

\subsubsection{Chaos}
Stiller and Radons also noticed that the model is chaotic
outside the domain of freezing,  i.e.~in most of
the $(\eta,J)$ phase plane, see Fig.~8 in Ref.~\onlinecite{StiRad98}.
Nevertheless a detailed analysis of the scaling of
chaos was not pursued. In the standard Kuramoto model
the Lyapunov exponent is known to scale quadratically with the
coupling constant \cite{popovych05}, but, to our knowledge,
this remains unknown for the DSR model.
We end noting that the volcano phase may perfectly coexist
with chaos, while freezing and chaotic dynamics are mutually exclusive.

\subsection{Methodologies}

The DSR can be analyzed following two main complementary strategies:
direct numerical simulations and self-consistent methods.
The advantages of each approach depends on the features under study.

The direct numerical integration of the model, Eq.~\eqref{eq.phasmod}, for fixed parameters and number of oscillators is the more straightforward approach. It was exclusively used in Refs.~\onlinecite{Dai92,SR00,Dai00,Pazo23},
and partly used in Refs.~\onlinecite{StiRad98,Nic25}.
Any dynamical quantity can be computed numerically,
though finite-size effects are unavoidable. Keeping these
effects sufficiently small entails a severe computational cost.
All the numerical simulations in this paper use this
straightforward method.
The reference value $\sigma=1$ is adopted, unless otherwise stated.
The numerical integration was performed with
a fourth-order Runge-Kutta algorithm of time step:
$$
\Delta t =\min(0.1,0.5/J)
$$
(or smaller in certain cases). The frequencies could be drawn at random or
deterministically sampled from $g(\omega)$ with
no effect in any of the results of this paper.

A second approach is to consider self-consistent
dynamic mean-field and cavity methods as in Refs.~\onlinecite{StiRad98,Pruser24a,Pruser24b,Kati24}.
They adopt the thermodynamic limit from the outset, reducing the infinite set of ODEs
to a self-consistent single-site problem, i.e., a
stochastic equation for one oscillator driven by self-consistent noises. This
self-consistent problem can be analytically solved perturbatively for weak
coupling \cite{Pruser24a,Pruser24b}. However, for general coupling,
the numerical self-consistence of noise is achieved iteratively.
To deal with the heterogeneity
of frequencies, noise
is obtained averaging over $M$ different
noisy oscillators
of frequencies $\{\omega_i\}_{i=1,\ldots,M}$,
introducing in this way $O(1/\sqrt{M})$ finite-size effects \cite{Kati24}.
Notice, nevertheless, that the scaling of the fluctuations is known in
contraposition to the less controllable $N$-dependent fluctuations of direct numerical
simulations.

The self-consistent method has shown its utility at computing the decay
of the order parameter\cite{StiRad98}, noise correlations\cite{Kati24} and the weak coupling dynamics\cite{Pruser24a,Pruser24b}.
However there is a caveat: Lyapunov exponents cannot be
computed using this approach; the evolution of an infinitesimal perturbation cannot be computed
self-consistently. This imply that to study the nature of chaos, our only viable alternative is direct numerical simulation.

We briefly comment on a third approach: the study of low-rank models.
These are models of the form of Eq.~\eqref{eq.phasmod} where
the coupling matrix $K$ is a random matrix with low rank \cite{bonilla93,Ottino18},
such that when the rank tends to infinity it converges to $K$ as defined
in Eq.~\eqref{K} \cite{Pazo23}.
If the rank is much smaller than the number of oscillators, those models can be solved through the Ott-Antonsen theory.
Unfortunately, the conclusions derived from low-rank models
do not generically apply to the full-rank model. The absence of algebraic
relaxation \cite{Ottino18} and the disparate values of critical asymmetry for the volcano transition \cite{Pazo23} illustrate the limitations of
low-rank models.

  \begin{figure*}[t]
	\includegraphics[width=\textwidth]{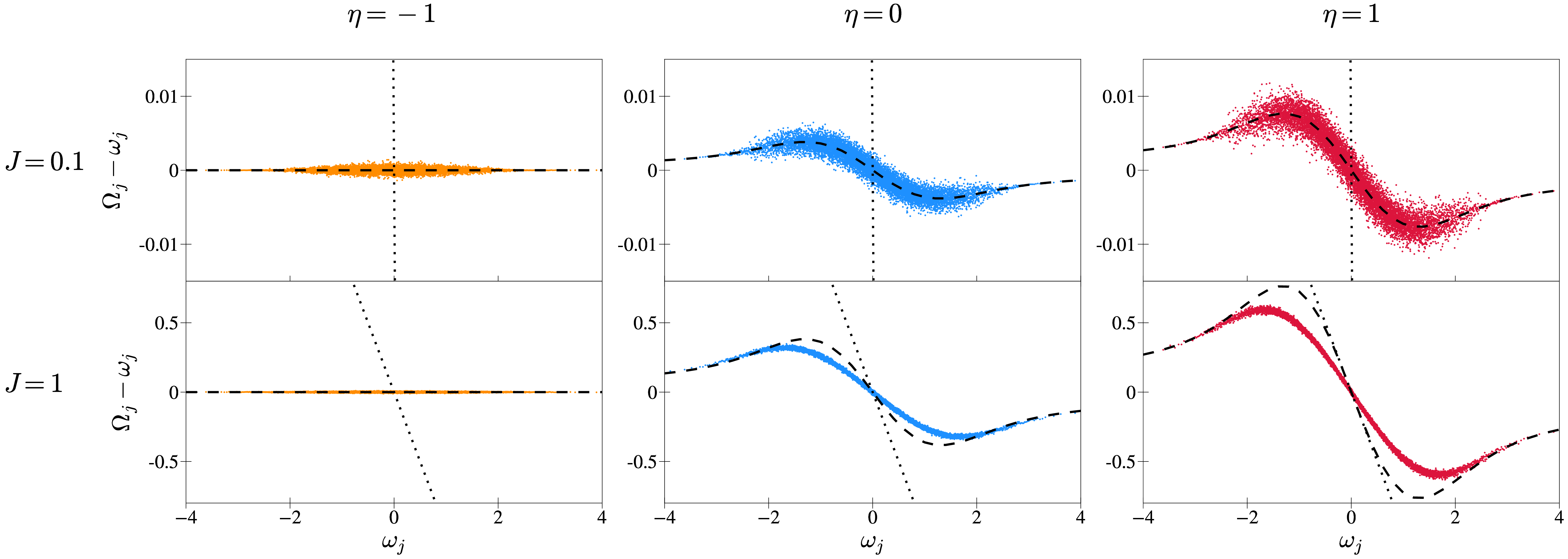}
	\caption
	{Difference between average and natural frequencies versus the natural frequency, for $N=10^4$ oscillators. In the upper (lower) row the coupling strength is fixed to $J=0.1$ ($J=1$). From left to right, the three columns
	correspond to $\eta=-1$, $0$ and $1$, respectively. The dashed line indicates the
	perturbative theoretical result in Eq.~\eqref{eq.freq}. The straight dotted line indicates
	the vanishing long-term average frequency ($\Omega_j=0$). }
	\label{fig.freq}
\end{figure*}

\section{Weak coupling} \label{sec.weakcoup}

We initially consider weak coupling. As a reference value, we
recall that the volcano transition occurs at $J_v\approx1.3 \sigma$
for $\eta=1$ (after proper rescaling of the recent numerical
determination in Ref.~\onlinecite{Pazo23}, see also \onlinecite{Pruser24a}).
Therefore it is reasonable to identify
the parameter region $J/\sigma\le1$, as
the weak coupling regime.

\subsection{Long-term average frequencies} \label{sec.freq}

 Repulsive and attractive
interactions are equally present in the DSR model. It
is therefore not obvious at first
sight if the oscillators should exhibit a tendency
to approach their actual frequencies or not.
The average frequencies
$\Omega_j=\langle\dot\theta_j\rangle_t$
are the first subject of our work.

Numerical simulations of Eq.~\eqref{eq.phasmod} were carried out with $N=10^4$ oscillators,
setting $\sigma=1$, and choosing
different combinations of $J=\{0.1,1\}$ and $\eta=\{-1,0,1\}$.
Multiple realizations of the coupling matrix and initial conditions were also implemented but no quantitative difference was noticeable. The average frequencies $\Omega_j$ were computed
over $t=10^5$ time units.
The results for the set of 6 combinations of parameters $\eta$ and $J$
are depicted in Fig.~\ref{fig.freq}. We found convenient to represent
the displacement of the frequency $\Omega_j-\omega_j$ against $\omega_j$. As each
panel has a different scale in the $y$-axis, we always include, as a reference,
a dotted straight line corresponding to $\Omega_j=0$ (the main frequency of
entrainment, if it existed).

The upper panels of Fig.~\ref{fig.freq} correspond to a
coupling constant $J=0.1$, a rather small value.
We see that, as $J$ is increased from zero,
the average frequencies move from $\omega_j$ towards 0,
save in the anti-reciprocal case ($\eta=-1$). In the latter case
the average frequencies
remain mostly unperturbed by the coupling $\Omega_j\approx\omega_j$.

The bottom panels of Fig.~\ref{fig.freq} show
the results for a larger coupling, $J=1$.
The $\Omega_j$ become even closer to 0,  specially those with small $|\omega_j|$.
 Again this frequency displacement heavily depends on $\eta$.
No trace of a frequency plateau is apparent, so oscillators are not entrained to any frequency.

By using the cavity method \cite{Pruser24a,Pruser24b}
we can obtain $\Omega_j$ as a perturbative expansion
in the small parameter $J$, see Appendix \ref{sec.app.cavity}.
At the first non-trivial order in $J$ we obtain:
\begin{equation}\label{eq.freq}
	\Omega_j=\omega_j-
	J^2\frac{1+\eta}{\sqrt{2}}\operatorname{D_w}(\omega_j/\sqrt{2})+O(J^4),
\end{equation}
where $\operatorname{D_w}(x)=\exp(-x^2)\int_0^x \exp(y^2) dy $ is the
so-called Dawson function.
We see that the leading correction to $\omega_j$ is
of order $J^2$. (The next order is $O(J^4)$ since  odd powers of $J$ are
absent due to the symmetry of the model
under $J\to-J$.) Moreover, the prefactor $(1+\eta)$ implies maximal
(null) frequency drift for $\eta=1$ ($-1$).
The function in Eq.~\eqref{eq.freq}
is represented by a dashed line in all panels of Fig.~1.
The agreement between Eq.~\eqref{eq.freq}
and the numerical results in Fig.~1 for $J=0.1$
is good.
For $J=1$ the numerical results exhibit a clear
deviation from the asymptotic formula \eqref{eq.freq} (save for $\eta=-1$),
although they are qualitatively similar.
The discrepancies at $J=1$ are not surprising ,
since Eq.~\eqref{eq.freq} is a perturbative result,
which already yields an
unphysical non-monotonic
dependence of $\Omega$ on $\omega$ for $J>\sqrt2$ ($\eta=1$).

We close this section noting that the numerical data do not
fall on a line, but form a ``cloud'' due to
the finiteness of the population.
The width of this cloud decreases with the number
of oscillators as $1/\sqrt{N}$ (not shown),
converging to a line in the thermodynamic limit.

\subsection{Scaling of the dissipation}\label{sec.dissip}

A theoretical analysis of the Lyapunov spectrum\cite{PPLYAP} of the DSR model is challenging.
Infinitesimal perturbations obey the following linear equations:
\begin{equation}
	\dot{\delta\theta}_j=\frac{J}{\sqrt{N}}\sum_{k=1}^{N}K_{jk}\cos(\theta_k-\theta_j)
	(\delta\theta_k-\delta\theta_j).
	\label{eq.infi}
\end{equation}
We can gauge the difficulties of achieving a theoretical result for the Lyapunov exponents,
if we take into account that the problem still remains  partly
solved in the standard Kuramoto model, see
Refs.~\onlinecite{popovych05,MGP2018,Carlu19}.

\begin{figure}
	\includegraphics[width=\linewidth]{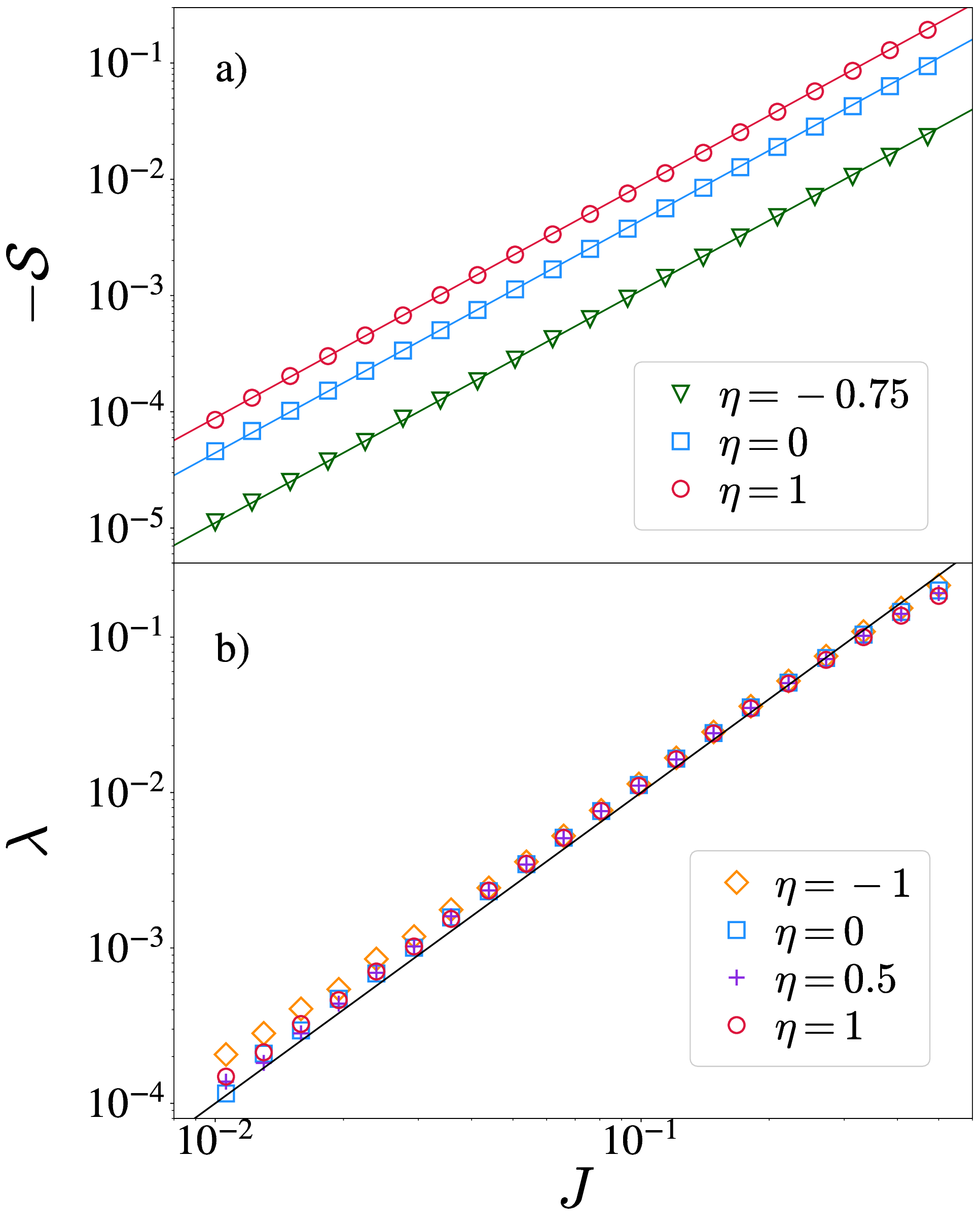}
	\caption
	{(a) Log-log plot of the dissipation ($- \cal S$) versus $J$ for different asymmetry parameters. The symbols correspond to numerical simulations with $N=5000$ oscillators, and the solid lines are the theoretical result in Eq.~\eqref{eq.diss}.
	(b) Largest Lyapunov exponent versus $J$ (in log-log scale) with
	$N=2000$, for different asymmetry parameters.
	The black solid line is $\lambda=J^2$.}
	\label{fig.lyap}
\end{figure}

Seeking a theoretical result, it is more productive to focus on
the average phase-space
volume contraction rate
per degree of freedom $\cal S$.
This quantity measures the dissipativeness of the model,
and equals the time average of the trace of the Jacobian matrix, see \eqref{eq.infi},
divided by $N$. Or, written  in mathematical form:
\begin{equation}\label{S}
	\mathcal{S}=-\frac{J}{N^{3/2}}\left<\sum_{jk}^{N}K_{jk}\cos(\theta_k-\theta_j) \right>_t.
\end{equation}
It is well known that $\cal S$ equals the sum of the Lyapunov exponents
\cite{PPLYAP} divided by $N$. We note that the finiteness of $\mathcal{S}$ is compatible with the presence of extensive chaos.

For an antisymmetric coupling matrix ($\eta=-1$) it
is immediate from the form of \eqref{S} that $\mathcal{S}=0$;
i.e.~the dynamics is conservative, as previously pointed out in
Refs.~\onlinecite{StiRad98,hanai24}.
For general $\eta$ we can
compute $\cal S$ perturbatively, by means of the cavity method:
\begin{equation}\label{eq.diss}
	\mathcal{S}=-(1+\eta)\frac{\sqrt{\pi}}{4}J^2+O(J^4)
\end{equation}
see appendix \ref{sec.app.cavity} for details. A log-log plot of this parabolic function is shown in Fig.~\ref{fig.lyap} (a)
for three selected values of $\eta$.
This figure also shows the results of direct numerical
determination of $\cal S$, via Eq.~\eqref{S} driven by Eq.~\eqref{eq.phasmod},
for $N=5000$. The agreement between theory and simulation is remarkably good.

\subsection{Scaling of the Lyapunov exponent} \label{sec.scale}

Here, we take a closer
look at the dependence of the Lyapunov exponent on the coupling
strength.
In Fig.~\ref{fig.lyap} (b), we depict the largest Lyapunov exponent $\lambda$
versus $J$ for $N=2000$ oscillators and different values of $\eta$. The growth
of the Lyapunov exponent is consistent with an
asymptotic quadratic law:
\begin{equation}
\lambda(J\ll1)\simeq c \, J^2 ,
\label{lJ2}
\end{equation}
where ``$\simeq$'' denotes equality after neglecting marginal contributions in $J$,
and $c$ is a constant of order 1.
Such a quadratic dependence on the coupling constant was already observed
in the Kuramoto model \cite{popovych05}, and other situations \cite{matias03}.
A striking peculiarity
of \eqref{lJ2} is the apparent independence of the constant $c$ on
parameter $\eta$. The small deviations for small $\lambda$ can be attributed
to numerical accuracy. The insensitivity of $\lambda$ to the value of $\eta$
is particularly intriguing, since, as seen above,
the average frequency and the dissipation do strongly depend on $\eta$.


The quadratic scaling of the Lyapunov exponent appears to be quite robust.
We computed the Lyapunov exponent with a coupling matrix whose elements were drawn from
a uniform distribution of unit variance.
We obtained the quadratic scaling \eqref{lJ2}, with the same
constant $c$ (not shown),
irrespective of the symmetry of the coupling matrix.
These empirical observations suggest the quadratic scaling in Eq.~\eqref{lJ2}
is the outcome of a general underlying mechanism at work in weakly coupled
oscillators.

\section{Identical Frequencies
(or Infinite Coupling)} \label{sec.infcoup}

Now we focus on the limit opposite  to the previous section.
We set $\sigma=0$, or equivalently $J\to \infty$, see Eqs.~\eqref{eq.phasmodrescaled} and \eqref{eq.phasmodrescaled2}.
This situation is already far from trivial.

\begin{figure}
	\includegraphics[width=\linewidth]{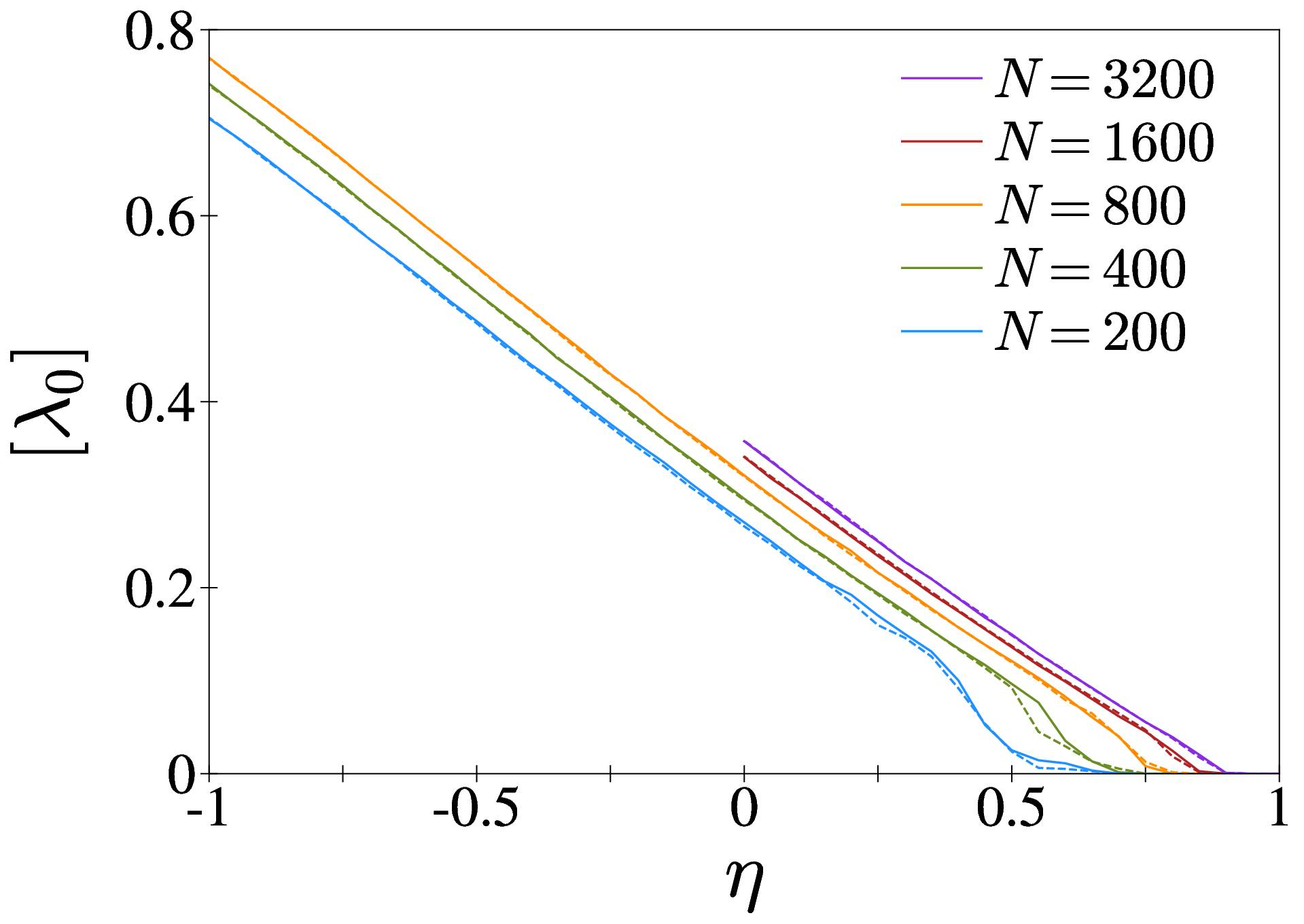}
	\caption
	{Largest Lyapunov exponent in the homogeneous  limit ($\sigma=0$),
	$\lambda_0$ versus $\eta$ for $N=200$, $400$, $800$, $1600$,  and $3200$ oscillators.
	The Lyapunov exponent has been averaged over 20 independent realizations, save for $N=3200$ with
	only 10 realizations. Solid (dashed) lines correspond to increasing (decreasing)
	$\eta$.}
	\label{fig.transition}
\end{figure}

Stiller and Radons concluded in Ref.~\onlinecite{StiRad98}
that the glassy state (freezing) persisted for $1\ge\eta>\eta_c$,
with $\eta_c(J\to\infty)\approx 0.8$.
Our goal is inferring
the behavior in the thermodynamic limit from the direct simulation
of increasingly larger systems. To study the $J\to\infty$ limit
it is convenient to set $\sigma=0$, and a finite $J$ value, e.g.~$J=1$.
In Fig.~\ref{fig.transition} we represent the numerically determined
Lyapunov exponent $\lambda_0(\eta)$.
We averaged over independent  realizations of the
coupling matrix, and denote the ensemble mean by $[\lambda_0]$.
Five different system sizes were considered.
To make our results more reliable, $[\lambda_0]$ was computed both
continuously increasing $\eta$ (solid lines) and decreasing $\eta$ (dashed lines),
see Appendix \ref{app.matrix} for details of the implementation.
It is apparent that chaos persists for increasingly larger values of
$\eta$ as  $N$ grows. For $N=3200$, the Lyapunov exponent
(averaged over 10 independent realizations of the coupling matrix)
approximately touches zero at $\eta_c\approx 0.9$,
which is clearly above the estimation by Stiller and Radons.
We conjecture that $\eta_c\to1$ as $N\to\infty$, i.e.~the glassy phase is
fragile against weak nonreciprocity.

Let us end this section mentioning that, by trivial rescaling,
a finite Lyapunov exponent $\lambda_0$
implies a linearly divergent Lyapunov exponent in the original
setup with fixed $\sigma=1$:
\begin{equation}
\lambda(J\gg1)\simeq \lambda_0(\eta) J .
\end{equation}


\section{Moderate and Strong coupling} \label{sec.modcoup}

We devote the remainder of this paper to explore the moderate and strong coupling regimes of the DSR model.
We first focus on reciprocal coupling, and
afterwards the picture is completed studying
the general non-reciprocal case.

\subsection{Reciprocal coupling} \label{sec.recip}

\subsubsection{Frequency entrainment and the volcano phase}

In the original work by Daido \cite{Dai92}
it is claimed that in the volcano phase, a portion of the oscillators became
``quasientrained''.
By this term, it is
meant frequency-entrained, with the phase differences displaying
a diffusive drift (i.e.~no phase locking).
This behavior is quite exotic for a deterministic system, and deserves a
deeper scrutiny. The perturbative result in Eq.~\eqref{eq.freq}
does not apply, and we resort to direct numerical simulations.

\begin{figure}
	\includegraphics[width=\linewidth]{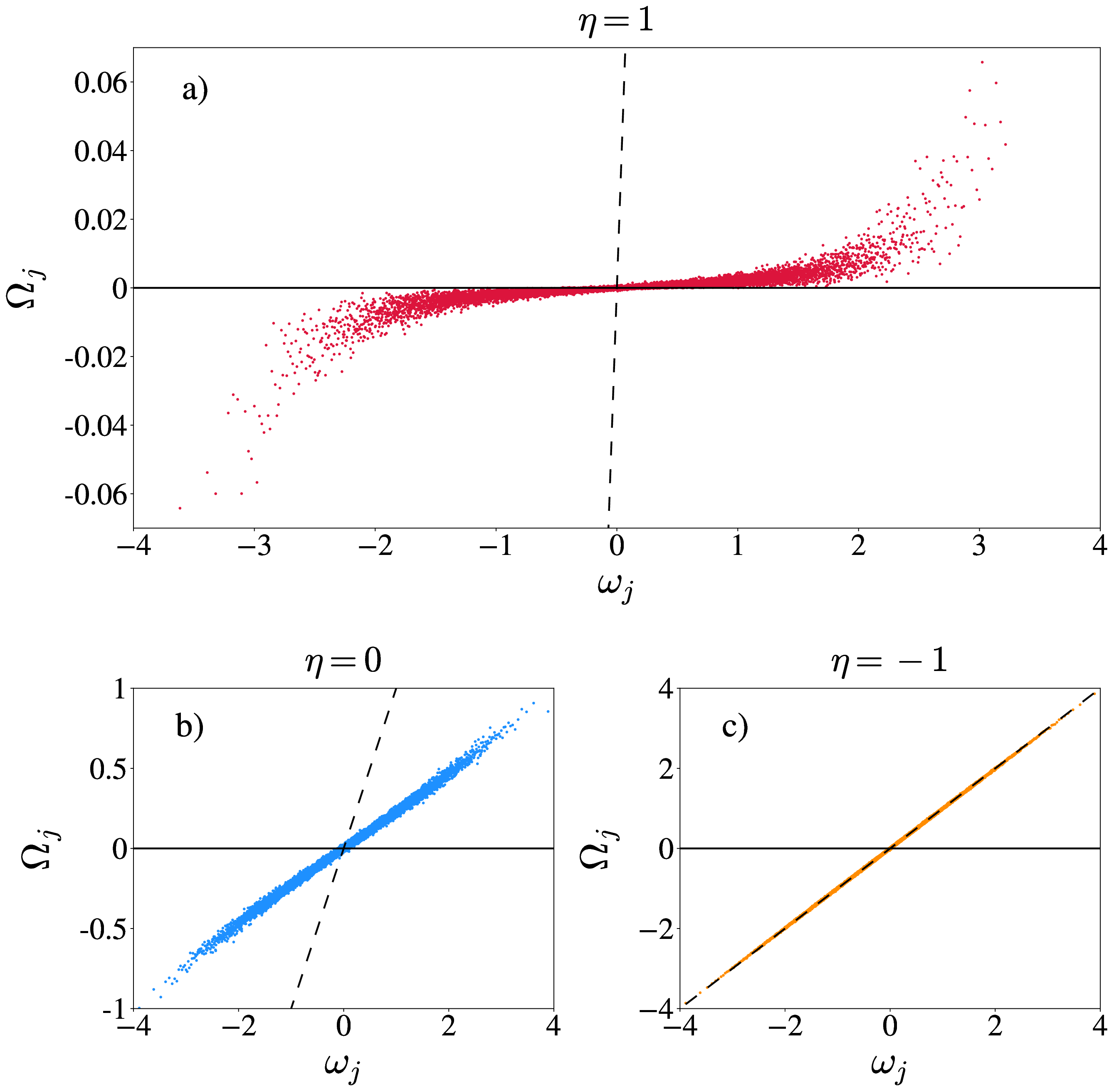}
	\caption
	{Long-term average frequencies of a population of $10^4$ oscillators
	with coupling constant $J=10$,
	and $\eta=1$, $0$ and $-1$ in panels (a), (b) and (c), respectively.
	The time averages were performed over $10^5$ t.u.
	The dashed straight line is $\Omega_j=\omega_j$.}
	\label{fig.freqlarge}
\end{figure}

In Fig.~\ref{fig.freqlarge}~(a), we depict the averaged frequencies for
$N=10^4$ oscillators (a fairly large value) and $J=10$,
a coupling strength well above the volcano transition. In the figure we observe oscillators with central natural frequencies settle into averaged frequencies close to zero, but a plateau cannot be distinguished.
This would suggest that, in the thermodynamic limit,
oscillators with different natural frequencies
display different averaged frequencies.
The diffusion of the phase difference observed in Ref.~\onlinecite{Dai92} is only tangible for short times.
In our simulations, we have observed that, for large enough integration times,
a ballistic evolution of the phase differences eventually prevails.
In other words, frequency entrainment does not emerge
with the volcano phase.
True phase diffusion would be only possible between oscillators
with identical natural frequencies. Because nonreciprocal coupling
is less favourable to synchronization, frequency entrainment is not
achieved neither for $\eta<1$, see Figs.~\ref{fig.freqlarge} (b) and (c). In fact, in the antireciprocal case $\eta=-1$, we observe the averaged frequencies coincide with the
natural ones.

\subsubsection{Multistability and freezing}

As already discussed in Sec.~\ref{sec.background},
in the infinite coupling or $\sigma=0$ limit the system displays
a conspicuous multistability of fixed points. Furthermore,
according to Stiller and Radons \cite{StiRad98}, the system also ``freezes'' for strong enough coupling. Is freezing
concomitant with multistability?
Are we even sure that freezing survives in the thermodynamic limit? Our next simulations intend to shed some light
on these questions.

\begin{figure}
	\includegraphics[width=\linewidth]{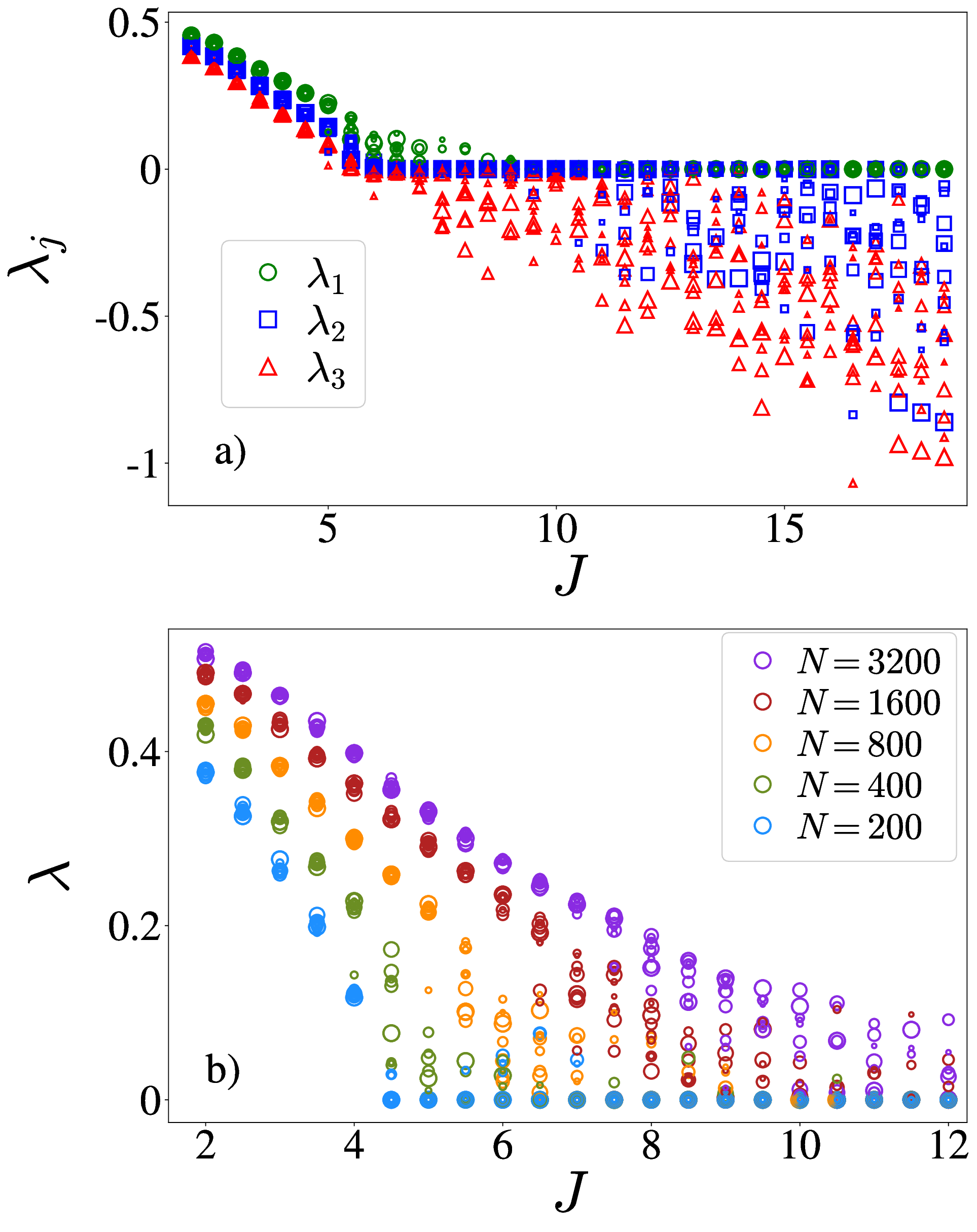}
	\caption{Lyapunov exponents for one instance of a symmetric coupling matrix $K$.
		(a) Three largest Lyapunov exponents $\lambda_1\ge\lambda_2\ge\lambda_3$ in green circles, blue squares and red triangles, respectively. The system size is $N=800$.
		Ten different initial oscillator phases were
		implemented for every value of $J$, and each
		symbol size corresponds to one of these runs.
		(b) Largest Lyapunov exponent $\lambda=\lambda_1$ for
		ten initial conditions, and five system sizes.
		}
	\label{fig.eta1}
\end{figure}

In Fig.~\ref{fig.eta1}(a) we represent the three largest Lyapunov
exponents as a function of the coupling strength for $N=800$ oscillators,
and a particular sample of the coupling matrix elements
and the natural frequencies. For every $J$ value, the Lyapunov exponents
were computed ten times starting with independent
random initial conditions
(i.e., ten ``replicas'' in the jargon of spin glasses).
For the largest values of $J$, we mostly obtain negative Lyapunov
exponents, save the first one, which is locked to 0 due to
the global rotational symmetry $\theta_j\to \theta_j+\alpha$.
The dependence of the negative Lyapunov exponents on the initial condition
indicates a multiplicity of resting  states
(nothing surprising in this context).
We also observe values of $J$ where two (or more) exponents are zero. This correspond to states in which most oscillators are entrained to a single frequency, while a few of them are entrained to another frequency, implying quasiperiodicity.

In Fig.~\ref{fig.eta1}(a) we may also observe that
chaotic and non-chaotic attractors coexist when $J$ is decreased below $J\approx8$.
Eventually, below $J\approx5$, we
infer that only one chaotic attractor survives,
as the same Lyapunov exponents are obtained
irrespective of the initial condition.

The attentive
reader may have noticed that the unentrained state
for $J=10$ in Fig.~\ref{fig.freqlarge}(a) is inconsistent with
the resting states found in Fig.~\ref{fig.eta1}(a).
This apparent discrepancy is due to the disparate population sizes
in both figures, $N=10^4$ vs. $N=800$. To clarify this issue,
we depict the largest Lyapunov exponent for different system sizes in Fig.~\ref{fig.eta1}(b).
Again, ten random initial conditions are set for every $J$ value.
The scenario of panel (a) applies to the other considered population sizes;
however, the disappearance of chaos is deferred as $N$ is increased.
From our simulations we cannot conclude if
there is a finite critical $J_c<\infty$ value
where chaos ceases in the thermodynamic limit.
The freezing transition described in Ref.~\onlinecite{StiRad98} was claimed
to occurs at $J_f\approx24$
(the value of $\sigma$ was not specified,
but we can presume $ \sigma=1$ as in our case).
Our simulations were not able to confirm
$J_c\approx J_f$, nor the mere existence of finite $J_c$ and $J_f$ values.
At this point, we can only assert that
the domain of chaos progressively grows as
the system size is increased. 
Similarly, in Ref.~\onlinecite{Nic25} the authors
note that, in order to observe the algebraic decay
of the order parameter $|Z|$,
coupling needed to be progressively  increased as $J\sim\sqrt{N}$.


\subsection{Non-reciprocal coupling} \label{sec.nonrec}

Finally, we explore the general non-reciprocal case
with non-small coupling. The lack of frequency entrainment
was already discussed above. Thus, we focus on the
chaotic dynamics.

\subsubsection{Pervasiveness of Chaos}

We now investigate the extension of chaos in the thermodynamic limit.
Specifically, we progressively increase the number of oscillators, and
determine the region where chaos exists.

\begin{figure}
	\includegraphics[width=\linewidth]{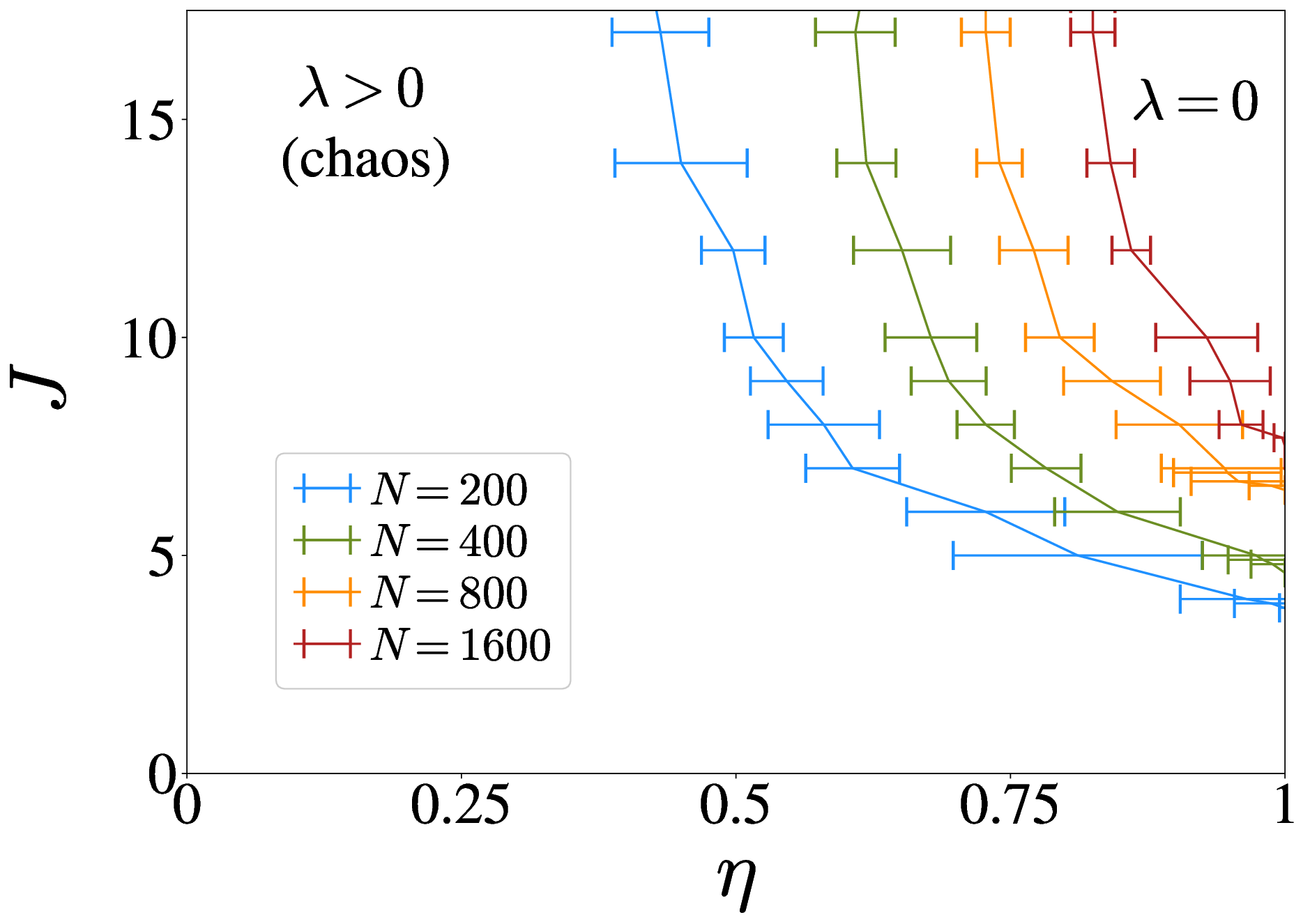}
	\caption
	{Partial phase diagram of Eq.~\eqref{eq.phasmod} for $N=200$, $400$, $800$ and $1600$, the solid lines indicate the mean critical value $\eta_c$ in which the Lyapunov exponent becomes smaller than $10^{-3}$. The error bar represent the dispersion among ten realizations of the coupling matrix.}
	\label{fig.phasediag}
\end{figure}

In Fig.~\ref{fig.phasediag} we show a partial phase diagram for
several population sizes. It is obtained computing the largest
Lyapunov exponent $\lambda$ for fixed values of $J$ and $N$, while, for each realization of the coupling matrix, we vary $\eta$. We iteratively detect
the value of $\eta$ at which $\lambda$ becomes smaller than
a threshold value $10^{-3}$.
This critical $\eta_c(J)$ is averaged over 10 realizations of the coupling matrix.
As can be observed, the region with vanishing Lyapunov exponent
shrinks as $N$ increases.
We interpret this result, combined with Fig.~\ref{fig.transition}, as an indication
that, in the thermodynamic limit, the system displays non-chaotic behavior
only for reciprocal coupling ($\eta=1$).

The pervasiveness of chaotic attractors when the system size grows is not new in systems with random interactions. In the Kuramoto-Sakaguchi model with random distributed phase lags, recently studied in Ref.~\onlinecite{PikBagn24}, increasing the number of oscillators translates into larger basin of attractions of chaotic attractors.

\subsubsection{Crossover between quadratic to linear growth of $\lambda(J)$}

\begin{figure}
	\includegraphics[width=\linewidth]{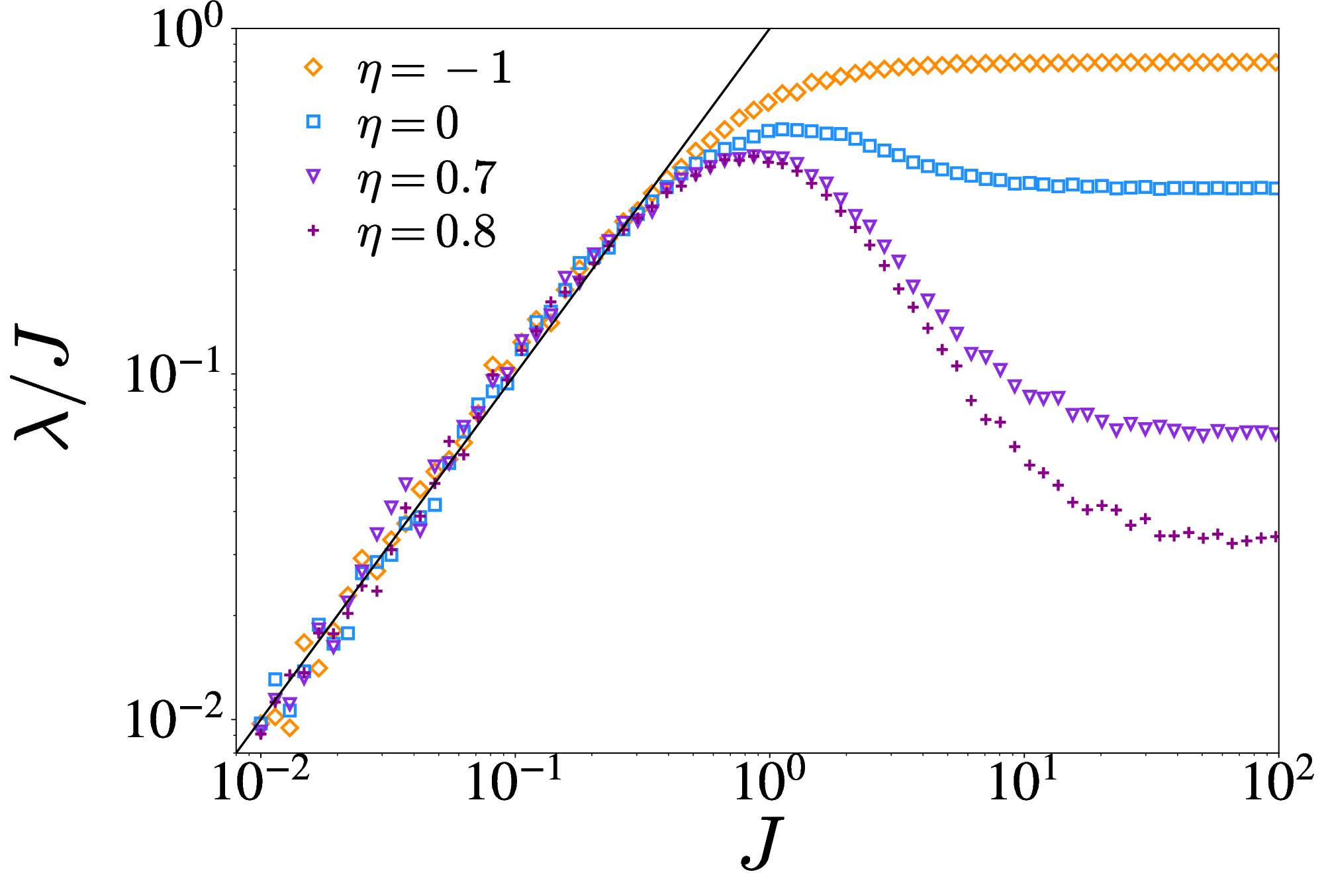}
	\caption
	{Largest Lyapunov exponent rescaled by $J$, $\lambda/J$ versus $J$ for $N=2000$ oscillators and different values of $\eta$. The black solid line is linear function,
	showing that the scaling of the exponents for small $J$ values
	is $\lambda\sim J^2$.}
	\label{fig.lyapt}
\end{figure}

Finally, we study the behavior of the
largest
Lyapunov exponent with $J$.
In Fig.~\ref{fig.lyapt}, we depict it rescaled by
$J$ for several values of the asymmetry
parameter $\eta$ ($N=2000$). For small $J$ we observe the quadratic scaling
discussed in Sec.~\ref{sec.weakcoup},
while for large values of $J$, the exponent grows linearly as expected
from the simulation in Sec.~\ref{sec.infcoup}.
The Lyapunov exponent exhibits an intricate crossover between
a quadratic law ($\lambda\sim J^2$) and
a linear one ($\lambda\sim J$) as $\eta$ approaches one. Somehow,
the system ``feels'' the lack of chaos for $\eta=1$ at $J\to\infty$.
The monotonicity of $\lambda(J)$ is lost at some point in the range
$0.7<\eta<0.8$.

%

\section{Conclusions} \label{sec.conc}

In this paper we have studied the dynamics of
the fully disordered Kuramoto model (or Daido-Stiller-Radons model\cite{Dai92,StiRad98})
of phase oscillators, Eq.~\eqref{eq.phasmod}. Our work
---via numerical simulations and theoretical analysis---
sheds light on the dynamics and chaotic properties of the system.

In the weak coupling limit we obtained some analytical results by means
of the cavity method. Specifically closed expressions for the long-term average frequencies
and the dissipation were derived using the dynamical cavity method.
Both quantities display a strong dependence on the reciprocity parameter $\eta$.
In contrast,
numerical simulations suggest that
the first Lyapunov exponent is asymptotically identical for all $\eta$ values.
Elucidating the underlying mechanism for this
is an interesting question left for future work.

In the infinite coupling limit, or the identical oscillators case,
the DSR model exhibits a glassy phase if the interactions are reciprocal.
Incorporating non-reciprocity causes the system to become chaotic.
Our simulations, see Fig.~\ref{fig.transition}, suggest that the threshold
$\eta$ value for chaos approaches 1 in the thermodynamic limit.

In the moderate coupling regime, we showed quasientrainment does not occur
if large enough integration times and population sizes are considered.
We also observed the importance of finite-size effects, since the critical
coupling where the system displays freezing strongly depends on $N$.
The exploration of the general non-reciprocal case evidences the pervasiveness
of chaos. A peculiar crossover between the quadratic and the linear scalings of the Lyapunov exponent $\lambda(J)$ is uncovered.
The value of the asymmetry parameter $\eta$
determines the shape of the mentioned crossover.

In sum, understanding the significance and the implications of
glassy (or quasi-glassy) phases in nonequilibrium systems,
from neuroscience to optics, constitutes a fascinating
research enterprise \cite{Far_beyond}.
In this context, the DSR model emerges as a basic prototypical model,
since self-sustained oscillators are pervasive in nature and technology.
The results presented in this paper may be deemed as limited, but we believe
they will pave the way for future progress. As usual in science,
our findings also give rise to interesting open problems.

\section*{Acknowledgements} 
We acknowledge support by
Grant No.~PID2021-125543NB-I00,
funded by MICIU/AEI/10.13039/501100011033 and by ERDF/EU.

\section*{DATA AVAILABILITY}
 
 The data that support the findings of this study are available
from the corresponding author upon reasonable request.

\appendix

\section{Dynamical-cavity calculation of the long-term average frequencies and the dissipation}\label{sec.app.cavity}

We sketch the dynamical-cavity calculation
used to obtain the
long-term
average frequency and the dissipation
in Eqs.~\eqref{eq.freq} and \eqref{eq.diss}, respectively.
The derivation is based on the results of Refs.~\onlinecite{Pruser24a,Pruser24b},
so we encourage the interested reader to check those references.

As a first step, it is useful to notice that both
quantities,
the shift of the averaged frequencies and dissipation, can be written
in terms of a local quantity $M_j$ related to the local field \eqref{eq.LF}:
\begin{equation}
	M_j=r_je^{i(\psi_j-\theta_j)} \label{mj}
\end{equation}
The real and imaginary parts, after
appropriate averaging, yield:
\begin{eqnarray}
	\Omega_j-\omega_j&=&J\langle \operatorname{Im}( M_j)\rangle_t , \label{oo}\\
	\mathcal{S}&=&-\frac{J}{N}\sum_{j=1}^N\langle \operatorname{Re} (M_j)\rangle_t. \label{sm}
\end{eqnarray}
Here $\langle f(t)\rangle_t=\lim\limits_{T\to\infty}\frac{1}{T}\int _0^Tf(t)dt$ is the time average.

The cavity method considers an ensemble of $N$ oscillators and adds another oscillator $\theta_0$ (and its corresponding matrix elements $K_{0k}$ and $K_{k0}$). Because the new elements are uncorrelated with the evolution of the $N$ previous oscillators, the order parameter $r_0e^{i\psi_0}$ can be computed by means of the central limit theorem. The evolution of the new oscillators is described (in the thermodynamic limit) by a stochastic equation subject to self-consistent noise\cite{Pruser24a,Pruser24b}:
\begin{multline}\label{Eq.Prüser}
	\partial_t\theta_0=\omega_0+\frac{J}{2i} \left(e^{-i\theta_0(t)} \xi(t)-e^{i\theta_0(t)} \xi^*(t)\right) \\-J^2 \eta \int_{0}^{t}dt' R(t,t')\sin(\theta_0(t)-\theta_0(t'))
\end{multline}
In this expression $\xi$ is a complex Gaussian noise, which accounts for the
thermodynamic limit of the local field $r_0e^{i\psi_0}$ in the absence of feedback
($\eta=0$).
The noise $\xi$ has zero mean and the autocorrelations of real and imaginary parts
are encoded by these two averages:
\begin{subequations} \label{eq.noise}
 \begin{eqnarray}
\langle\xi(t)\xi^*(t')\rangle_\xi&=&2C(t,t'), \\
\langle\xi(t)\xi(t')\rangle_\xi&=&0.
\end{eqnarray}
\end{subequations}
It is simple to calculate $C$ at the lowest order in $J${\cite{Pruser24a,Pruser24b}
for a Gaussian distribution of frequencies:
\begin{equation}
	C(t,t')=\frac{1}{2}e^{-\frac{(t-t')^2}{2}}+O(J^2)
\end{equation}
The last term in Eq.~\eqref{Eq.Prüser} accounts
for the feedback effect of the original ensemble (the cavity)
on the new oscillator after being perturbed by it. At the lowest order\cite{Pruser24a,Pruser24b}:
\begin{equation}
	R(t,t')=H(t-t')\frac{1}{2}e^{-\frac{(t-t')^2}{2}}+O(J^2)
\end{equation}
where $H(x)$ is the Heaviside step function.

Hence the local order parameter for oscillator $\theta_0$ at time $t$ (Eq.~(35) and (36) in Ref.~\onlinecite{Pruser24b}) is:
\begin{equation}
	r_0e^{i\psi_0}=\xi(t)+\eta J\int_{0}^{t}e^{i\theta_0(t')}R(t,t') dt' \label{lf0}
\end{equation}

The dynamics of $\theta_0$ in Eq.~\eqref{Eq.Prüser} can be perturbatively described
by expanding the solution in powers of $J$:
$\theta_0(t)=\theta_0^0+J\theta_0^1+\cdots$.
This yields $\theta_0^0=\omega_0 t$ and
\begin{equation}\label{eq.solth1}
	\theta_0^1(t)=\frac{1}{2i} \int_{0}^{t}e^{-i\omega_0 t'} \xi(t')-e^{i\omega_0 t'}\xi^*(t')dt'.
\end{equation}
If this solution is plugged into Eq.~\eqref{lf0} we get:
\begin{equation}
	r_0e^{i\psi_0}=\xi\nonumber+\eta J e^{i\omega_0 t}I(t)+O(J^2)
\end{equation}
where we have defined
\begin{eqnarray}
	I(t)&=&e^{-i\omega_0 t}\frac{1}{2}\int_{0}^{t}e^{i\omega_0t'}e^{-\frac{(t-t')^2}{2}} dt'
\end{eqnarray}

With the previous results, we can turn our view to the
relevant quantity in Eq.~\eqref{mj}.
Up to linear order in $J$ we get
\begin{equation}
	\langle M_0\rangle_t\simeq \langle \xi e^{-i\omega_0 t}-iJ \xi e^{-i\omega_0 t}\theta_0^1(t)+\eta J I(t)\rangle_t \label{M0t0}
\end{equation}
As uncertainties vanish in the thermodynamic limit
(self-averagingness), time averages are
independent of the realization of the coupling matrix
or equivalently the realization of the noise,
that is $\langle\rangle_{t}=\langle\rangle_{t,\xi}$. Thus
we evaluate the first two terms in the right hand-side
of Eq.~\eqref{M0t0}:
\begin{eqnarray}
	\langle \xi(t) e^{-i\omega_0 t}\rangle_t&=&\langle \langle\xi(t)\rangle_\xi e^{-i\omega_0 t}\rangle_t=0 \nonumber\\
		-i\langle \xi(t) e^{-i\omega_0\nonumber t}\theta_0^1(t)\rangle_t&=&-i \langle e^{-i\omega_0\nonumber t}\langle\xi(t)\theta_0^1(t)\rangle_\xi \rangle_t\\
		&=&\left<\int_{0}^{t}e^{i\omega_0(t'-t)} C(t,t')dt' \right>_t \nonumber\\ &\simeq& \langle I(t)\nonumber\rangle_t
\end{eqnarray}
In the last equality we have used Eqs.~\eqref{eq.solth1} and \eqref{eq.noise}.
Then Eq.~\eqref{M0t0} simplifies to:
\begin{equation}
	\langle M_0\rangle_t=(1+\eta)J\langle I(t)\rangle_t+O(J^3)
\label{M0t}
\end{equation}
It can be shown that terms of order $J^2$ vanish since an odd number of Gaussian noises need to be averaged in its derivation. The time average of $I(t)$ can be computed
\begin{equation}
	\langle I(t)\rangle_t\simeq\sqrt{\frac{\pi}{8}}e^{-\omega_0^2/2}-\frac{i}{\sqrt{2}}\operatorname{D_w}(\omega_0/\sqrt{2})
\end{equation}
Inserting this expression into Eq.~\eqref{M0t}, and
using Eq.~\eqref{oo} it is straightforward to
obtain Eq.~\eqref{eq.freq} in the main text.

To compute $\mathcal{S}$ from Eq.~\eqref{sm}
we need to average over all oscillators.
In the thermodynamic limit this is equivalent to average
over frequencies, obtaining:
\begin{equation}
	\mathcal{S}\simeq-(1+\eta)J^2\sqrt{\frac{\pi}{8}}\int_{-\infty}^{\infty}g(\omega)e^{-\omega^2/2}d\omega ,\end{equation}
The evaluation of the integral immediately yields Eq.~\eqref{eq.diss} in the main text.

\section{Generation of the coupling matrix} \label{app.matrix}
In this appendix we describe how we generate a coupling matrix with cross correlation between symmetric elements given by $\langle K_{jk}K_{kj}\rangle =\eta$. We first build a symmetric matrix $S$ and an antisymmetric matrix $A$ whose elements are drawn from a zero-mean unit-variance normal distribution. The coupling matrix $K$
is computed as:
\begin{equation}
	K=\sqrt{\frac{1+\eta}{2}} S +\sqrt{\frac{1-\eta}{2}} A.
\end{equation}
By changing $\eta$ we go continuously from an antisymmetric to a symmetric coupling matrix, as done for Fig.~\ref{fig.transition} and Fig.~\ref{fig.phasediag}.

\end{document}